\documentclass[prl,twocolumn,amsmath,amssymb,aps]{revtex4-2}

\usepackage{comment}
\usepackage{xcolor}
\usepackage{graphicx}
\usepackage{dcolumn}
\usepackage{bm}
\usepackage{hyperref}
\usepackage{kantlipsum}
\hypersetup{pdfborder=0 0 0,colorlinks=true,citecolor=blue,linkcolor=blue}

\begin{document}

\title{Josephson effect as signature of electron-hole superfluidity \\
in bilayers of van der Waals heterostructures}
\author{Filippo Pascucci$^{1,2}$, Sara Conti,$^3$ David Neilson,$^3$ Jacques Tempere,$^2$ and Andrea Perali$^{1*}$}
\affiliation{$^1$Supernano Laboratory, School of Pharmacy, University of Camerino, 62032 Camerino (MC), Italy}
\affiliation{$^2$TQC, University of Antwerp, Universiteitsplein 1, 2610 Antwerp, Belgium}
\affiliation{$^3$CMT, Dept.\ of Physics, University of Antwerp, Groenenborgerlaan 171, 2020 Antwerp, Belgium}

\date{\today}

\begin{abstract}
We investigate a Josephson junction in an electron-hole superfluid in a double layer TMD heterostructure. 
Observation of a critical tunneling current is a clear signature of superfluidity. 
In addition, we find the BCS-BEC crossover physics in the narrow barrier region controls the critical current across the entire system. 
The corresponding critical velocity, which is measurable in this system, has a maximum when the excitations pass from bosonic to fermionic. 
Remarkably, this occurs for the density at the boundary of the BEC to BCS-BEC crossover regime determined from the condensate fraction. 
This provides, for the first time in a semiconductor system, an experimental way to determine the position of this boundary.
\end{abstract}

\maketitle

Recent experimental reports of quantum condensation with bound pairs of spatially separated electrons and holes in double layers of graphene \cite{Burg2018} or Transition Metal Dichalcogenide (TMD) van der Waals heterostructures \cite{Wang2019}, have created a flurry of new experimental and theoretical investigations.
Spatially separating the electrons and holes prevents them from recombining, opening the way to a low temperature stable, long-lived superfluid \cite{Lozovik1976}.
By varying the equal carrier densities in the two layers using metal gates, the superfluid can be tuned from the strongly coupled Bose-Einstein condensation (BEC) regime of compact boson-like particles to an intermediate coupling regime, and to the Bardeen-Cooper-Schrieffer (BCS) regime with more extended fermionic pairs \cite{Conti2017,LopezRios2018}. 

From an application viewpoint, a supercurrent in an electron-hole superfluid can carry a particle flow without dissipation when the electron and hole layers are independently contacted in a counterflow configuration \cite{Su2008}. 
This directly leads to applications in dissipationless solid-state electronics \cite{Su2008,Nandi2012}. 

The neutrality of the electron-hole pairs creates a challenge to unambiguously identify the fluid as a superfluid \cite{Su2008}. 
Observing a non-dissipative counterflow current \cite{Tutuc2004,Liu2022} by itself is not sufficient to claim superfluidity.
The conventional criterion to identify superconductivity, the combination of perfect conduction and the Meissner effect (perfect diamagnetism) cannot be used with neutral pairs. 
Coulomb drag resistance and enhancement of interlayer tunneling conductance \cite{Burg2018,Wang2019} with accompanying electroluminescence measurements \cite{Wang2019}, may sometimes be used to indirectly identify the superfluidity. 

Here we propose that the Josephson effect \cite{Josephson1962, Zenker2015} 
can provide an unambiguous signal of the electron-hole superfluidity. 
In a Josephson junction, two superfluids are separated by a thin potential barrier, and a phase difference between the superfluids leads to a steady current flow.
Observation of a dissipationless current through the barrier when there is no driving potential present is regarded as an optimal direct experimental way to confirm the existence of the single amplitude and phase of the macroscopic wave-function that characterizes a quantum condensed state \cite{Ancilotto2009,Spuntarelli2007}.

We find that the maximum value of the dissipationless current exhibits a notable sensitivity to the bosonic or fermionic nature of the low-lying excitations of the superfluid state in the barrier region. 
We further show that this sensitivity can be exploited to identify and distinguish in the barrier region, 
the BEC regime of bosonic-like pairs from the weakly bound fermion pairs of the BCS–BEC crossover and BCS regimes \cite{Pascucci2020}.

\begin{figure}[bt]
\includegraphics[trim=2.7cm 0.0cm 1.0cm 0.0cm, clip=true, 
width=\columnwidth]{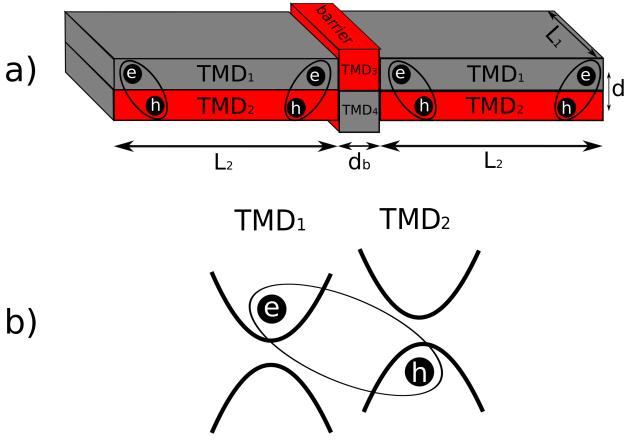}
\caption{(a) Schematic of Josephson junction with the different layers labelled TMD$_1$--TMD$_4$. $d$ is layer separation, $d_b$ the barrier thickness, and $L_1$ and $L_2$ the transverse and longitudinal layer lengths.
Electron-hole pairs are shown.
(b) Energy band alignments at the type-II TMD$_1$/TMD$_2$ interface. }
\label{Fig1}
\end{figure}

The Josephson junction can be fabricated using a combination of lateral stitching and vertical stacking of TMD heterostructures \cite{Taghinejad2019}, as represented in Fig.\ \ref{Fig1}(a). 
A vertical stacking of two different TMD monolayers, TMD$_1$ and TMD$_2$, is separated by a thin barrier made of two different undoped TMD monolayers, TMD$_3$ and TMD$_4$.
The potential barrier height is determined by the difference in energy of the conduction (valence) bands in the doped TMDs and in the undoped TMDs of the barrier.
The critical Josephson current can be measured in a counter-flow configuration \cite{Su2008}, using the  method of Anderson \cite{Anderson1963} and Shapiro \cite{Shapiro1963}.

We select the TMD$_1$ and TMD$_2$ of the vertical stacking to have type-II interface, with the edges of the conduction and valence band at different energies (Fig.\ \ref{Fig1}(b)). 
This keeps the electrons and holes spatially separate without the need for an insulating barrier. We consider a single-band with one interaction channel. 

The coupled BCS  mean-field equations for the superfluid gap $\Delta_k$ and density $n$, remain a good approximation in the BCS-BEC crossover and BEC regimes at zero temperature \cite{Conti2020b,Perali2013}, 
\begin{eqnarray}
\Delta_k &=& -\frac{1}{L_1 L_2} \sum_{k'}\mathcal{F}_{1,2}V^{sc}_{k-k'}\frac{\Delta_{k'}}{2E_{k'}}\ ,
\label{gap}\\
n &=& \frac{g_s}{L_1 L_2} \sum_k \frac{1}{2} \left(1-\frac{\varepsilon_{k}-\mu_s}{E_k}\right)\ .
\label{number}
\end{eqnarray}
We set the electron and hole densities $n$ equal, and take equal effective masses in the TMD single-particle parabolic bands $\varepsilon_k$. 
The excitation energy $E_{k}=\sqrt{\xi_k^2+\Delta_k^2}$, with $\xi_k=\varepsilon_k-\mu_s$ and $\mu_s$ the single-particle chemical potential. The form factor $\mathcal{F}_{1,2}$ accounts for the overlap between the single-particle wave functions in TMD$_1$ and TMD$_2$. The spin degeneracy is $g_s=2$, and $L_1$ and $L_2$ the transverse and longitudinal layer lengths.    

In Eq.\ (\ref{gap}), $V^{sc}_{k-k'}$ is the effective self-consistent screened Coulomb attraction between electrons and holes.  
Because of the long-range nature of Coulomb interactions, screening plays a crucial role here \cite{Neilson2014}. We use the expression in Ref.\ [\citenum{Perali2013}], which  self-consistently takes into account the weakening of the screening in the presence of a superfluid energy gap,
\begin{equation}
V^{sc}_{\mathbf{q}}=\frac{V^{D}_\mathbf{q}} 
{1-2V^{S}_\mathbf{q}(\Pi_n(\mathbf{q})+\Pi_a(\mathbf{q}))
+\mathcal{A}_\mathbf{q}\mathcal{B}_\mathbf{q}} \ .
\label{Eq:VeffSF} 
\end{equation}
$V^{S}_\mathbf{q}$ and $V^{D}_\mathbf{q}$ are the bare Coulomb interactions 
within a layer and between  layers, respectively.   
$\Pi_n(\mathbf{q})$ and $\Pi_a(\mathbf{q})$ are the normal and anomalous polarizabilities.
For brevity, in Eq.\ (\ref{Eq:VeffSF}) we write
$\mathcal{A}_\mathbf{q}=(V^{S}_\mathbf{q})^2-(V^{D}_\mathbf{q})^2$ and 
$\mathcal{B}_\mathbf{q}=\Pi^2_n(\mathbf{q})-\Pi_a^2(\mathbf{q})$.

The critical velocity of the superfluid is given by the Landau criterion \cite{Landau1941},
\begin{equation}
v_c=\min_{k}\frac{\mathcal{E}_k}{\hbar k}\ .
\label{Landaucrit}
\end{equation}
There are two types of excitation energy $\mathcal{E}_k$ in this system, 
Anderson-Bogoliubov modes associated with  bosonic 
behavior of the pairs, and the fermionic modes associated with pair-breaking excitations.

In the bosonic excitation branch, $\mathcal{E}_k$ is given by the dispersion relation 
$\mathcal{E}_k=\hbar k c_{s\!f}$ \cite{Bogoliubov1947}, 
where $c_{s\!f} =\sqrt{{\mu_{s\!f}}/{2m}}$ is the speed of sound,
with superfluid chemical potential $\mu_{s\!f}=2\mu_s+\varepsilon_B$, and   
 $\varepsilon_B$ the binding energy of a single electron-hole pair. 
 From Eq.\ (\ref{Landaucrit}), the critical velocity for  bosonic excitations is thus the speed of sound, 
\begin{eqnarray}
v_c^{(BEC)}=c_{s\!f} =\sqrt{\frac{\mu_{s\!f}}{2m}}\ .
\label{cs}
\end{eqnarray}

Instead, for single-particle fermionic excitations $\mathcal{E}_k=E_k$, and the critical velocity is the pair-breaking (p-b) velocity \cite{Combescot2006},
\begin{equation}
v_c^{(p-b)}= \min_{k} \frac{\sqrt{(\varepsilon_k-\mu_s)^2+\Delta_k^2}}{\hbar k}. \ 
\label{vc(p-b)}
\end{equation}
is numerically evaluated for given values of $\mu_s$ and $\Delta_k$  
to determine the value of $k=k_{\text{min}}$ that minimises Eq.\ (\ref{vc(p-b)}).  
As the density is increased and the superfluid regimes are scanned from BEC to BCS-BEC crossover to BCS, 
the critical velocity should switch from $v_c^{(BEC)}$ to $v_c^{(p-b)}$, whichever is the lesser. 

We consider only sufficiently wide barriers for the Thomas-Fermi local approximation to be valid, 
$d_b > \xi$, where $ \xi={\hbar}/{mv_c}$ is the superfluid coherence length \cite{Combescot2006,Giorgini2008}. 
In the barrier region, the single-particle chemical potential $\mu^b_s$ 
and the superfluid chemical potential $\mu^b_{s\!f}$ will be reduced compared with their values outside the barrier,
\begin{eqnarray}
    \mu^b_s &=& \mu_s-V_0 \  ,
\nonumber \\
\mu^b_{s\!f} &=&\mu_{s\!f}-2V_0 = 2(\mu_{s}-V_0)+\varepsilon_B \ .
    \label{musb}
\end{eqnarray}
Using $\mu^b_s$ in Eqs.\ (\ref{gap}) and (\ref{number}) gives
the superfluid gap $\Delta_k^b$ and the density $n^b$ inside the barrier region. 
For a low rectangular potential barrier, 
$V_0 < \mu_{s\!f}/2$, current can flow over the barrier across a superfluid region 
in the barrier with density $n^b < n$.  
For high potential barriers, $V_0\geq\mu_{s\!f}/2$, $\mu^b_{s}<-\varepsilon_B/2$ and $n^b=0$, so the current across the barrier is given purely by quantum tunneling of the electron-hole pairs \cite{Meier2001,Zaccanti2019}.  

\begin{figure}[bt]
\includegraphics[trim=0.0cm 0.0cm 0.0cm 0.0cm, clip=true, width=\columnwidth]{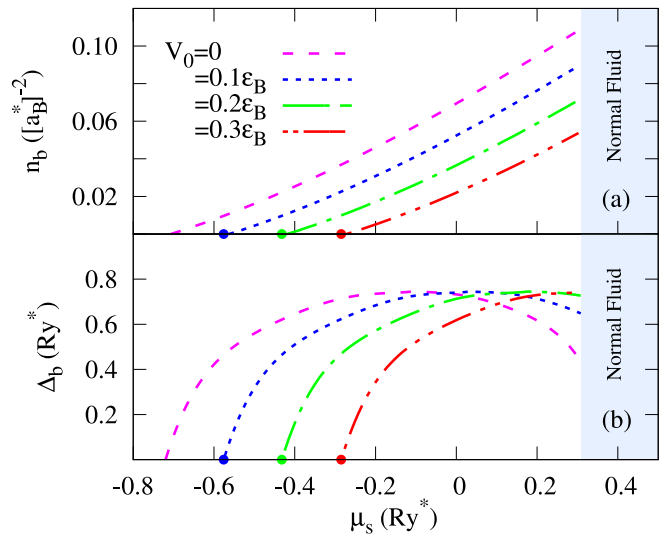}
\caption{(a) Density $n^b$ inside the barrier as a function of the single-particle chemical potential $\mu_s$ outside barrier.  $V_0$ is barrier height with $\varepsilon_B=1.42$ Ry$^*$. In shaded area, strong screening suppresses the superfluidity. (b) Superfluid gap $\Delta^b$ inside the barrier. }
\label{Fig2}
\end{figure}

Figure \ref{Fig2}(a) shows for different barrier heights $V_0$ in the low barrier regime, the superfluid density in the barrier $n^b$ as a function of the $\mu_s$ outside the barrier. 
We use effective Rydbergs Ry$^*$ for the energy scale, and effective Bohr radii a$_B^*$ for the length scale. Table \ref{barrier_height} gives values for Ry$^*$ and a$_B^*$ for different TMD heterostructures to connect with experimental results. 
For $d=0.6$nm, the typical layer separation of a TMD type-II interface, $\varepsilon_B=1.42$ Ry$^*$. 
The color-coded dots indicate the value $\mu_s=-\varepsilon_B/2+V_0$ below which $n_b$ is zero.  
The BEC regime is characterised by negative values of $\mu_{s}$.
As $\mu_s$ increases and becomes positive, the system enters the BCS-BEC crossover regime, but $\mu_s$ remains well below the Fermi energy.  
Only in the weak-coupled BCS limit would $\mu_s$ approach the Fermi energy.  
However, for sufficiently large $\mu_s$, strong screening of 
the electron-hole pair interaction in the superfluid 
outside the barrier region suppresses the superfluidity \cite{Perali2013} 
(the shaded regions in Fig.\ \ref{Fig2}).  
The $V_0=0$ curve in Fig.\ \ref{Fig2}(a) gives as a reference the superfluid density $n$ 
in the absence of a barrier. 
This reaches a maximum at the value $\mu_s=0.31$ Ry$^*$ for density $n=n^b=0.105 (a_B^*)^{-2}$. This defines the onset density $n_0$ for the superfluidity.

Figure \ref{Fig2}(b) shows the maximum of the superfluid gap $\Delta^b$ inside the barrier as a function of $\mu_s$.  
The curve for $V_0=0$, also gives the maximum 
of the superfluid gap $\Delta$ in the 
superfluid regions outside any barrier and, like $n$, 
this gap vanishes when $\mu_s$ reaches $-\varepsilon_B/2$. 
For $V_0 > 0$, $\Delta^b$ for the barrier vanishes at the same value of $\mu_s$ at which $n^b$ vanishes.

The barrier height $V_0$ can be varied by suitable material choice of TMDs 
with Type-II interfaces.  
Table \ref{barrier_height} gives examples.  
As a final example, WS$_2|$WSe$_2$---MoSe$_2|$WSe$_2$, the barrier is inserted only in 
the electron monolayer, a configuration which may be more straightforward
to fabricate.
\begin{table}[t]
\renewcommand{\arraystretch}{1.2}
\resizebox{\columnwidth}{!}{%
\begin{tabular}{lcl|cccc}
\hline
\hline
TMD$_1|$TMD$_2$&---&TMD$_3|$TMD$_4$&$a_B^*$[nm]&\ \ Ry$^*$[meV] &\ \ \ $\varepsilon_B$[meV]\ \ \  \ \ &$V_0[\varepsilon_B]$\\ 
\hline
MoS$_2|$MoSe$_2$ &---&MoSe$_2|$WSe$_2$ &1.3 &100 &140 &0.04\\
MoS$_2|$WSe$_2$  &---&MoTe$_2|$WTe$_2$ &1.7 &77  &108 &0.05\\
WS$_2|$WSe$_2$  &---&MoTe$_2|$MoTe$_2$ &1.8 &71  &99  &0.10\\
MoS$_2|$WS$_2$   &---&WS$_2|$MoSe$_2$  &1.7 &77  &108 &0.20\\
MoSe$_2|$MoTe$_2$&---&MoS$_2|$MoSe$_2$ &1.1 &116 &162 &0.33\\
MoS$_2|$MoSe$_2$ &---&WTe$_2|$WSe$_2$  &1.3 &100 &140 &0.71\\
\hline
WS$_2|$WSe$_2$   &---&MoSe$_2|$WSe$_2$ &1.9 &71  &99  &0.33\\
\hline
\end{tabular}
}
\caption{Material parameters and barrier heights $V_0$.}
\label{barrier_height}
\end{table}

For low potential barriers $V_0<\mu_{s\!f}/2$  there is significant superfluid flow over the barrier.
The critical current in the barrier region is, 
\begin{equation}
I_c^b = n^b L_2 v_c^b \ .
\label{IcbV}
\end{equation}
The critical velocity $v_c^b$ in the barrier is the lesser of $v_c^{b(BEC)}$ and $v_c^{b(p-b)}$, obtained from Eqs.\ (\ref{cs}) and (\ref{vc(p-b)}) with $\mu^b_{s\!f}$ and $\Delta^b_k$.

For high potential barriers, $V_0>\mu_{s\!f}/2$, quantum tunnelling of the electron-hole pairs 
through the barrier determines the critical current,
\begin{equation}
\hbar I_c^b=n_c\, t_{s\!f}(\mu_{s\!f})\ L_1L_2\ .
\label{IcVh}
\end{equation}
$n_c={\cal{C}} n$ is the density of the superfluid condensate, with ${\cal{C}}$ the condensate fraction of the superfluid state. The transfer matrix element, 
\begin{equation}
t_{s\!f}(\mu_{s\!f})=f(V_0/\mu_{s\!f})\frac{\mu_{s\!f}}{k_{\mu_{s\!f}}L_1} \text{e}^{-k_{\mu_{s\!f}}d_b}
\label{tcc}\ ,
\end{equation}
is the probability for an electron-hole pair to tunnel across the barrier, where $k_{\mu_{s\!f}}^{-1}={\hbar}/\sqrt{2m(V_0-\mu_{s\!f})}$ is the wave-function decay length in the barrier, and 
\begin{equation}
f(V_0/\mu_{s\!f})=\left[1-\frac{V_0}{\mu_{s\!f}}-\sqrt{\left(\frac{V_0}{\mu_{s\!f}}\right)^2-1}\, \right]^2\ .\\
\end{equation}
So for $V_0>\mu_{s\!f}/2$, the final expression for the critical current is,
\begin{equation}
I_c^b=\frac{2n_0\, \mu_{s\!f}f(V_0/\mu_{s\!f}) \text{e}^{-k_{\mu_{s\!f}}d_b}\ L_2}{\hbar k_{\mu_{s\!f}}}
\label{Ichb}\ .
\end{equation}
\begin{figure}[bt]
\includegraphics[trim=0.0cm 0.0cm 0.0cm 0.0cm, clip=true,width=\columnwidth]{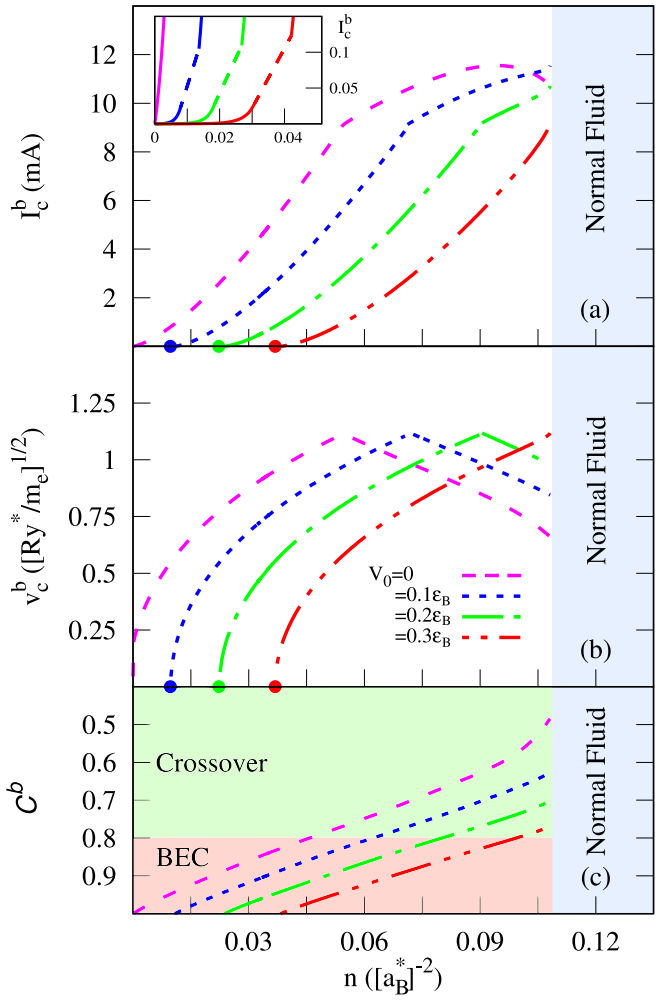}
\caption{(a) Critical current $I_{c}$ in the barrier for barrier height $V_0$, as a function of density $n$.  Inset: details of the critical current at very low densities; dashed lines interpolate high-barrier and low barrier results. (b) Critical velocity $v_c^b$ in barrier. (c) Condensate fraction ${\cal{C}}^b$ in the barrier.} 
\label{Fig3}
\end{figure}
\begin{figure}[bt]
\includegraphics[trim=0.0cm 0.0cm 0.0cm 0.0cm, clip=true,width=\columnwidth]{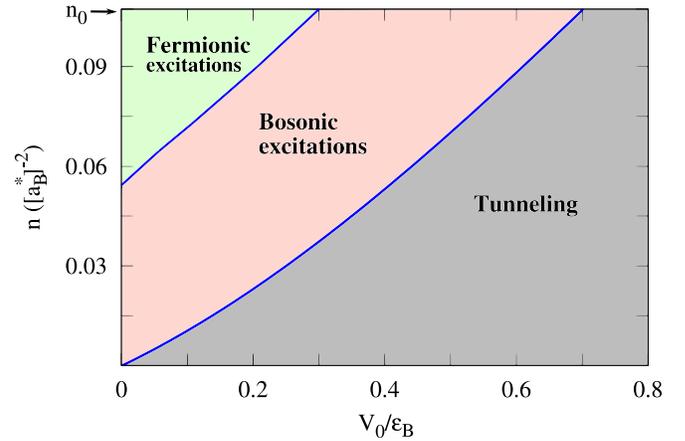}
\caption{Driving mechanisms for the Josephson critical current at different barrier heights $V_0$.
}
\label{Fig4}
\end{figure}

Figure \ref{Fig3}(a) shows the critical current in the barrier $I_c^b$ as a function of the density $n$ outside the barrier.
The colored dots again mark the value of $n$ at which $n_b$ drops to zero and the switch occurs from predominantly flow over to tunneling. 
The inset shows in detail the critical current in the tunneling region. 
This is seen to connect smoothly (dashed lines) with the critical current in the flow over region. We recall that the existence of a non-zero tunneling current in this region is accepted as a clear signature of superfluidity.  
The flattening of $I_c^b$ at high densities, reflects the drop in $\Delta^b_k$ from the strong screening (Fig.\ \ref{Fig2}(b)).
  
We note that $I_c^b$ is everywhere less than the critical current outside the barrier $I_c=nL_2v_c$, shown by the $V_0=0$ curve. For this reason, the overall critical current in the system is given by $I_c^b$. Thus the BCS-BEC crossover physics in the barrier region controls the transport properties of the entire device.

Figure \ref{Fig3}(b) shows the critical velocity $v_c^b$ across the barrier.
The maxima in $v_c^b$ result from the switch from Anderson-Bogoliubov bosonic excitations to single-particle fermionic excitations, $v_c^{(p-b)}$ increasing with density while $v_c^{(BEC)}$ decreases with density \cite{Miller2007}. 
As expected, the position of the maxima are sensitive to the barrier height.
Remarkably, Fig.\ \ref{Fig3}(c) shows that the maximum of $v_c^b$ for each value of $V_0$ matches the density at which the condensate fraction ${\cal{C}}^b=0.8$, the position of the BEC to BCS-BEC crossover boundary \cite{Guidini2014}. 

It is an attractive concept and relevant for experiments, that the switchover from bosonic excitations to single-particle fermionic excitations lines up with the BEC and BCS-BEC crossover regime boundary. 
In contrast to the condensate fraction which is not observable, the critical velocity $v_c^b=I_c^b/n$ is a directly experimentally measurable quantity in these electron-hole Josephson devices: $I_c^b=I_c$, the overall critical current of the system, and the density $n$ is precisely controlled by gate potentials. 
This remarkable result provides a way of experimentally locating the BCS-BEC crossover boundary.

In Fig.\ \ref{Fig4}, we show the nature of the driving mechanisms of $I_c^b$ for different $V_0$ and $n$. The density $n$ is capped at the superfluid onset density, $n_0$.
For very small $V_0$, as we increase density, we go from bosonic excitations to fermionic pair-breaking excitations. On increasing $V_0$, a region of tunneling of electron-hole pairs appears at small $n$.
When $V_0>0.3\varepsilon_B$, strong screening preempts $v_c^b$ from reaching the maximum, so there are no pair-breaking fermionic excitations.  
For high potential barriers, $V_0>0.7\varepsilon_B$, there are no bosonic excitations, and only tunneling through the barrier remains.

We have demonstrated that measurements of the critical current across a Josephson-junction barrier can yield significant 
additional information on electron-hole superfluid properties in a double-layer TMD heterostructure.  
The barrier can be fabricated and its height 
adjusted by suitable combinations of  TMD layers. 
The additional information is as follows.  
(i) The existence of a Josephson effect below a 
critical tunneling current is {\it per se} a direct signature of superfluidity.
We note that this could be used to distinguish between a phase of excitons in a normal or superfluid state.  
Up to now, this has required painstaking analysis to merge Coulomb drag resistance and 
counterflow experimental data \cite{Liu2022}. 
(ii) For low barriers, the crossover physics in the barrier region controls the  transport properties of the entire device.
(iii) One can experimentally observe the maximum of the critical velocity at the density where excitations switch from bosonic to fermionic, the density in this system controlling the coupling strength.
This maximum can be used to identify the boundary separating the BEC and BCS-BEC crossover regimes of the electron-hole superfluidity, and in fact, remarkably, the density at the maximum matches the density at which the condensate fraction passes through $0.8$. 

The work was partially supported by the projects G061820N, G060820N, G0H1122N and by the Flemish Science Foundation (FWO-Vl). 

\begin{small}
$^*$Correspondence to: Andrea.Perali@unicam.it
\end{small}

\end{document}